# Charting the Exciton-Polariton Landscape of WSe$_2$ Thin Flakes by Cathodoluminescence Spectroscopy

Masoud Taleb[1], Fatemeh Davoodi[1], Florian K. Diekmann[1], Kai Rossnagel[1,2], Nahid Talebi[1*]

[1]Institute for Experimental and Applied Physics, Kiel University, 24098 Kiel, Germany

[2]Ruprecht Haensel Laboratory, Deutsches Elektronen-Synchrotron DESY, 22607 Hamburg, Germany

E-mail: talebi@physik.uni-kiel.de

**Semiconducting transition-metal dichalcogenides (TMDCs) provide a fascinating discovery platform for strong light-matter interaction effects in the visible spectrum at ambient conditions. While most of the work has focused on hybridizing excitons with resonant photonic modes of external mirrors, cavities, or nanostructures, intriguingly, TMDC flakes of sub-wavelength thickness can themselves act as nanocavities. Here, we determine the optical response of such freestanding planar waveguides of WSe$_2$, by means of cathodoluminescence spectroscopy. We reveal strong exciton-photon interaction effects that foster long-range propagating exciton-polaritons and enable direct imaging of the energy transfer dynamics originating from cavity-like Fabry-Pérot resonances. Furthermore, confinement effects due to discontinuities in the flakes are demonstrated as an efficient means to tailor mode energies, spin-momentum couplings, and the exciton-photon coupling strength, as well as to promote photon-mediated exciton-exciton interactions. Our combined experimental and theoretical results provide a deeper understanding of exciton-photon self-hybridization in semiconducting TMDCs and may pave the way to optoelectronic nanocircuits exploiting exciton-photon interaction beyond the routinely employed two-oscillator coupling effects.**

The phenomenon of spontaneous collective coherence in condensed matter quantum systems, as most notably manifested by Bose-Einstein condensation, continues to be a subject of increasing interest[1-5]. Whereas temperature is the typical control knob to drive a system of collective excitations to its ground state, quasiparticle density in solid-state systems, such as the density of plasmons[6] or excitons[7,8], has recently emerged as a versatile control parameter[9], toward high-temperature condensation. In general, excitons are promising for condensation due to their light mass and bosonic nature. More specifically, excitons in semiconducting group VI transition-metal dichalcogenides (TMDCs), are advantageous because of their exceptionally high binding energies of a few 100 meV giving rise to stable and robust exciton excitations at room temperature[10-12]. In addition, the high oscillator strengths associated with the excitonic resonances result in exciton lifetimes longer than 100 fs, enabling effective light-mater coupling[13]. For these reasons, TMDCs are now a particularly fruitful platform for the observation of strong coupling effects between excitons and photons[14] including the formation of exciton-polariton quasiparticles with effective masses reduced by several orders of magnitude[15].

In order to realize strong exciton-photon couplings in practice, high-quality microcavities[16-19], plasmonic nanoantennas[6], or plasmon polaritons[20,21] typically need to be used. The effective interaction between cavity photons and excitons generally promotes the creation of exciton-polaritons, with spontaneous coherence enabling coherent energy transfer and the possible use in optoelectronic devices that, e.g., combine ultrafast optical routing with electronics functionalities[22,23]. However, the application of microcavities often involves complicated manufacturing processes limiting achievable coupling strengths, and microcavities also typically sustain only a limited operational bandwidth and are usually large in size, which hinders their use in hybrid electronic-optics nanocircuits. Plasmons, on the other hand, themselves suffer from large dissipative losses[24]. Therefore, simple effective platforms for strong exciton-photon coupling are in principle needed. Recently, thin TMDC waveguides were investigated along these lines by optical microspectroscopy and scanning near-field microscopy, where exciton-photon anticrossing behavior in individual flakes was observed[25,26] and exciton-polariton propagation was directly imaged using monochromatic photons[27,28], respectively. Yet, the underlying coupling mechanisms remain largely unclear because a spectroscopic investigation at high spatial resolution and at a broad energy range is missing. Moreover, spectroscopy techniques utilizing spontaneous interactions, in contrast with laser-based techniques, could be used to unambiguously unravel the mechanism of spontaneous coherence compared to lasing.

Here, we investigate the spontaneous optical properties of $WSe_2$ thin flakes by means of cathodoluminescence (CL) spectroscopy[29,30]. Using an electron beam-based probe, we rule out the possibility of lasing as the underlying coherence mechanism. Transversal one-dimensional optical confinement within the thin film and the propagation of the optical waves along the longitudinal orientation result in strong exciton-photon coupling as manifested in a Rabi energy splitting of 0.24 eV (corresponding to a wavelength splitting of approximately 110 nm) and the formation of exciton-polaritons whose spatial coherence and propagation mechanisms are investigated at high spatial resolution. Moreover, dispersion control of the optical modes via film thickness, stress engineering, two-dimensional and ultimately three-dimensional confinement is used to tailor the exciton-photon coupling strength. Finally, multi-oscillator coupling physics is explored on the basis of photon-mediated exciton-exciton couplings between A and B excitons, i.e., polariton-polariton interactions.

## Results

**Cathodoluminescence spectroscopy of $WSe_2$ waveguides.** Electron microscopy has continuously been advanced as a probe of both structural and optical properties of materials, with arguably the highest spatial, temporal and energy resolutions[31]. Regarding the optical properties, typically either electron energy-loss spectroscopy (EELS)[32] or CL spectroscopy[30,33] are used. CL spectroscopy, in particular, has proven powerful in analyzing exciton and bandgap excitations in semiconductors and defects[34]. It has only recently been revolutionized to probe not only incoherent processes, but also spontaneous coherence, particularly of surface plasmon polaritons[35], localized plasmons[36], and optical modes of photonic crystals[37,38]. However, to the best of our knowledge, CL spectroscopy has not yet been applied to probe exciton-polariton coherence and spatial correlations, which is partially due to the competing incoherent and coherent radiation channels in semiconductors[39]. Below, we will show that CL can probe exciton polaritons and Rabi splitting in atomically flat single-crystalline TMDC flakes of sub-wavelength thickness.

Specifically, we investigate the optical response of 80 nm thick flakes of the prototypical exciton-active TMDC semiconductor $WSe_2$, using 30-keV electrons in a field-emission SEM system equipped with a parabolic mirror and optical detectors (see Methods Section for details). Excitons are excited upon electron irradiation and exciton-photon interaction, exciton-polariton formation, as well as the propagation dynamics are probed by CL spectroscopy (Fig. 1a). Whereas in semiconductor optics weak exciton-photon interaction is omnipresent, strong interaction requires standing-wave-like patterns of light with increased interaction time. This is normally realized by optical cavities. However, the atomically flat interfaces of a material can also act as mirrors exploiting total internal reflection of the light. Thus, photonic modes of a slab waveguide[40], propagating within the *xy*-plane and forming standing-wave patterns along the transverse *z*-axis, facilitate both strong exciton-photon interactions and exciton-polariton transfer dynamics (Fig. 1b). The corresponding $WSe_2$ slab waveguides were fabricated by liquid exfoliation and placed on top of holey carbon transmission electron microscopy (TEM) grids (Fig. 1c) or carbon discs.

**Theoretical description of strong exciton-photon coupling in $WSe_2$ thin films.** The coupled harmonic oscillators, underlying the strong exciton-photon interactions, are modelled by virtue of decomposing the optical responses (for example the permittivity of the material) versus photons captured in a slab waveguide and the Lorentzian responses associated with certain excitons; here the A and B excitons in $WSe_2$. To better account for inhomogeneous broadening, the generalized model reported elsewhere for modelling the permittivity of the gold[41] is adopted here as well (see Supplementary Note 1 and Supplementary Fig. S1 for details). A dielectric slab waveguide can support optical modes of the so-called transverse electric ($TE_x$) and transverse magnetic ($TM_x$) polarizations[40]. Within the thickness ranges considered here, the fundamental $TE_x$ and $TM_x$ modes are both excited; however, only the $TM_x$ mode can strongly interact with the excitons and be revealed by electron spectroscopy techniques. The latter is due to the fact that electrons can only interact with those photonic modes supporting an electric-field component oriented along the electron trajectory[42]. The photonic modes supported by a dielectric slab waveguide with $\varepsilon_{r\infty} = 15$ can strongly interact with the A excitons in the WSe$_2$ thin films, leading to an energy splitting and level repulsion (Fig. 1d): a phenomenon often referred to as anticrossing[43]. The resulting polaritonic branches, referred to as Lower Polaritons (LP) and Upper Polaritons (UP), emerge in the energy ranges below and above the A exciton energy, respectively, and exhibit rather similar group velocities. For a slab waveguide with the thickness of $d = 80\ nm$, the energy gap opening is on the order of $\delta E = 0.24$ eV, $\delta \lambda = 110$ nm. Whereas at all thicknesses of $d > 60\ nm$, the photonic modes can strongly interact with the A excitons,

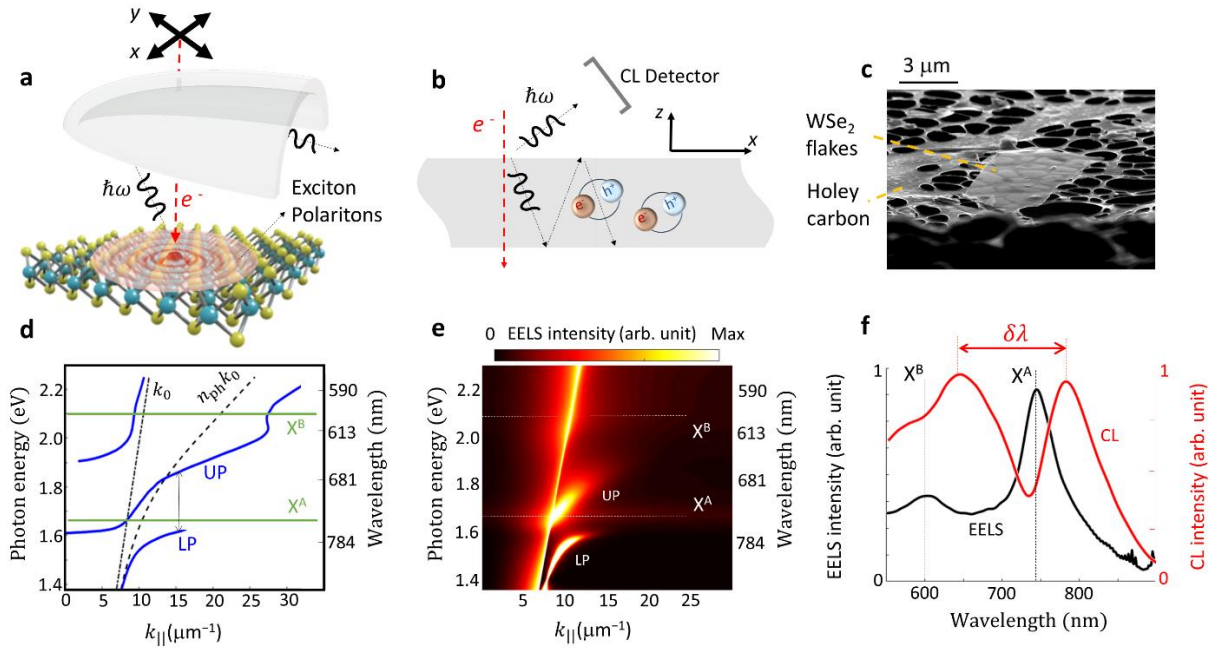

**Figure 1. Strong exciton-photon interactions in WSe$_2$ slabs.** (a) Schematic of the experimental setup where exciton-polaritons in a WSe$_2$ thin film are excited by an electron beam and probed using cathodoluminescence spectroscopy. (b) Schematic of a thin film supporting quasi-propagating optical excitations that are confined in the transverse direction and can strongly interact with excitons. (c) High-magnification image of thin-film WSe$_2$ flakes as deposited on a holy carbon grid. (d) Schematic dispersion diagram where strong interaction of quasi-propagating photonic modes with A excitons causes an energy splitting and anticrossing in the energy-momentum relation of the emergent exciton-polaritons. Exciton energies and photon dispersions

are indicated by solid and dashed lines, respectively. (e) Simulated energy-momentum EELS map, where the anticrossing is resolved for a 80-nm thick WSe$_2$ film. (f) Comparison of EELS and CL spectra. The EELS spectrum is governed by exciton absorption peaks. In contrast, strong-coupling effects and a resulting energy splitting are apparent in the CL spectrum (see Methods section). In (e) and (f) exciton energies are indicated by dotted lines.

we do observe only weak interactions with the higher-energy B excitons. This is due to the excitation of higher-order and yet radiative photonic channels with their dispersion diagram positioned inside the light cone, and the radiative nature of the B excitons as well, that does not allow for enhanced exciton-photon interactions. Exciton B has 3 times faster relaxation rates compared to exciton A, as can be understood by comparing $\gamma_A = 29743.43 \, \text{nm}$ ($\tau_A \simeq 99.2 \, \text{fs}$), to $\gamma_B = 10056.85 \, \text{nm}$ ($\tau_B \simeq 33.5 \, \text{fs}$) (see Supplementary Note 1 and Supplementary Fig. 1). This effect results in much shorter interaction times between photons and the B exciton, and hence the interaction between photonic modes and the B exciton is weaker. This behavior does not depend on the thickness of the slab waveguide, but indeed lateral confinements along the *y*-axis can lead to the modification of the dispersion of the photonic modes, thus allowing for observing strong interactions with both A and B excitons, as will be shown later.

A momentum-resolved electron energy-loss spectroscopy (MREELS) map, calculated for an electron with a kinetic energy of 30 keV interacting with a slab waveguide of thickness $d = 80 \, nm$, can unravel the dispersion of propagating optical modes (Fig. 1e), and demonstrates the same energy splitting and anticrossing as our model analysis predicted (see Supplementary Note 2). This demonstrates in principle the ability of swift electrons with kinetic energy as low as 30 keV (and even lower; see Supplementary Fig. 2) to efficiently couple to exciton-polaritons in WSe$_2$ thin films and to probe their coherent dynamics. However, the required momentum resolution at lower energy-loss ranges imposes a challenge for current instruments, since the polariton-induced angular recoil experienced by the electrons covers only a few microradians for relativistic electron beams of the sort used in transmission electron microscopes. Moreover, since both radiative and loss channels contribute to EELS [44], probing coherent exciton-photon interactions and analysis of transport properties are less straightforward using EELS compared to CL spectroscopy (Fig. 1f). In fact, the signal integrated over all the momentum components reveals that exciton absorption peaks are the dominant contributions to the overall electron energy-loss signal, for the thicknesses of interest of our thin films. In contrast, the observation of a dip – instead of a peak – at the A exciton wavelength in the CL spectra as well as a wavelength splitting as large as 110 nm demonstrates the possibility of using CL spectroscopy for probing exciton-polaritons in thin TMDC flakes. Moreover, propagating polaritons in infinite thin films normally do not contribute to the radiation channels and their dispersion line is positioned outside the light cone. Whereas the EELS signal constitutes propagating modes, the dip in the CL signal represents a depletion of energy due to the excitation of propagating modes. Interestingly, within the wavelength ranges of 749 nm $< \lambda <$ 774 nm and $\lambda <$ 652 nm, the LP branch and higher-order radiative polaritons contribute to the radiation continuum (Fig. 1d), whereas the photonic mode (dashed line in Fig. 1d) does not. Hence, the formation of the LP mode and the strong coupling effect result in two CL peaks, in contrast to the exciton absorption peak apparent in the EELS signal.

A deeper understanding of the level repulsion phenomenon in quantum electrodynamics systems, and the resulting Rabi oscillations, demands a more rigorous treatment of light-matter coupling. Here, we comply with a simple semi-classical model that perfectly captures the main mechanisms, defining the coupling between the electric field ($\vec{E}$) of the photonic modes and the macroscopic exciton polarization ($\vec{P}$) as[13]

$$i\dot{\vec{P}} = \left(\omega_{\text{ex}} - i\gamma\right)\vec{P} + g\vec{E}$$
$$i\dot{\vec{E}} = \left(\omega_{\text{ph}} - i\kappa\right)\vec{E} + g\vec{P} \tag{1}$$

where $\omega_{\text{ex}}$ and $\omega_{\text{ph}}$ are the angular frequencies of the exciton and photon oscillators, $\gamma$ and $\kappa$ are their damping constants, $g$ is the coupling strength, and time derivative is indicated by a dot on top. The eigenfrequencies of the coupled system of equations in Eq. (1) are derived by assuming time-harmonic responses ($P = P_0 \exp(i\omega t)$ and $E = E_0 \exp(i\omega t)$) and are given by

$$\omega_{\pm} = \frac{1}{2}\left(\omega_{ph}(k) + \omega_{ex} - i(\gamma + \kappa)\right) \pm \frac{1}{2}\sqrt{\Delta}, \tag{2}$$

with

$$\Delta = \left(\omega_{ph}(k) - \omega_{ex}\right)^2 + 2\Gamma\omega_{ph}(k)\omega_{ex} - (\gamma - \kappa)^2 + i2\left(\omega_{ph}(k) - \omega_{ex}\right)(\gamma - \kappa), \tag{3}$$

and $\Gamma = 2g^2\left(\omega_{ph}\omega_{ex}\right)^{-1}$. Equation (2) defines two solutions for the system associated with the LP and UP branches. Going beyond previous works, we allow for nonzero damping terms, thus the complex-valued parameter $\Delta$ can be used to model both Rabi oscillation and its damping. Intriguingly, for the resonance condition ($\omega_{ph}(k) = \omega_{ex}$), the damping of the Rabi oscillation is only determined by the damping of its underlying oscillators as $\text{Im}\{\omega_\pm\} = -(\gamma + \kappa)/2$. In contrast, detuning of the energy ($\hbar\delta\omega = \hbar\omega_+ - \hbar\omega_-$) can cause another mechanism for the relaxation of the Rabi oscillation. We have assumed a dispersion-dependent model for the photonic mode, allowing for efficient modeling of the anticrossing behavior similar to the complete treatment in Fig. 1d. Using the extracted parameters for the A exciton energy and photonic modes ($\omega_{ex} = 1.65\,\text{eV}$, $\gamma = 0.0417\,\text{eV}$, and $\kappa = 0$; see Supplementary Note 1), and under resonance condition ($\omega_{ph}(k) = \omega_{ex}$), we indeed observe an energy splitting $\Omega = \omega_+ - \omega_-$ of 0.24 eV for a coupling strength of $g = 121$ meV. A slightly larger coupling efficiency of $g = 163$ meV has been reported for WS$_2$ nanotube waveguides, which offer a more confined modal configuration and hence stronger exciton-photon interactions[45]. Interaction of excitons in WS$_2$ gratings with plasmons was reported to offer larger coupling efficiencies, on the order of $g = 205$ meV [46].

**Experimental probing of exciton-polaritons using CL spectroscopy.** In the experiments, we first look at the CL response of a WSe$_2$ flake with a thickness of 80 nm (Fig. 2). CL spectra, acquired at selected electron impact positions, demonstrate a wavelength splitting on the order of 120 nm, comparable to the predicted values, and a dip at the A exciton wavelength of 751 nm (Fig. 2c). This splitting is fully reproduced in the calculated CL spectra (Fig. 2g). Upon scanning the electron beam vertically to the edge of the flake, we observe several maxima with their relative distance depending on the wavelength (as indicated by the dashed arrows in Fig. 2b), and a dip at the A exciton energy splitting the excited polaritons into energy intervals above and below the exciton A energy associated with UP and LP branches, respectively. The energy of the photonic mode depends on the thickness of the slab waveguide. Hence, the exciton-photon coupling strength and the observed energy splitting strongly depend on the thickness of the slab waveguide (Fig. 2d). For a slab waveguide at a thickness of 65 nm, the splitting in the wavelength is on the order of $\Delta\lambda = 250$ nm, whereas by increasing the thickness to 80 nm, the wavelength splitting becomes $\Delta\lambda = 120$ nm. The increase in the wavelength splitting is understood by generating more resonant conditions between photons and excitons (Supplementary Fig. 4). Nevertheless, for larger thicknesses, the excitation of higher-order modes and the Cherenkov radiation result in exciton-photon interactions as well (Supplementary Fig. 6).

When comparing to the results of numerical calculations (using the finite-difference time-domain method, see Methods section), we indeed observe peaks at the predicted spectral and spatial positions in the two-dimensional CL spectrum. However, the maximum-minimum contrast, in the experimental data is less pronounced, which we attribute to structural imperfections, potentially caused by our liquid exfoliation and transfer methods, as well as excitation of incoherent channels, such as non-radiative electron-hole pairs. The occurrence of spatial interference fringes is well known in the formation of localized plasmonic modes in nanoscopic and mesoscopic plasmonic nanoantennas, such as nanorods[47] and micro-platelets[48]. Based on these similarities between propagating surface plasmon excitations[49-51] and the interference fringes observed in Fig. 2, we first hypothesize that reflection from the apex of the triangular flake might cause the interference fringes (Fig. 2e). The fringe periods would then be observable as a result of the constructive near-field interference and the ability of CL to probe such near-field patterns. For this to happen, constructive interference between the excited and reflected exciton-polaritons from the apex would imply $2\beta_{EP}L + \delta\varphi = 2m\pi$, where $\beta_{EP}$ is the phase constant of the surface polaritons, $L$ is the distance between the electron and the apex, and $\delta\varphi$ is the phase shift due to reflection from the apex. Thus, the period of the fringes would correspond to approximately one half of the wavelength of the excited polaritons. A comparison between the spatial interference fringes and the calculated exciton-polariton dispersion (Fig. 1d) rules out this hypothesis (since the resulting phase constant will be much smaller than the calculated phase constant of the $TM_x$ mode) as a possible reason for the observation of the interference fringes, and points to differences between EELS and CL, as the former is a probe of the near-field excitations, and the latter is a measure of far-field contributions. Our second hypothesis, thus, assumes that the observed interferences reflect interferences between the transition radiation and the scattering of exciton-polaritons from anomalies or edges, implying a far-field constructive interference pattern in the form of $\beta_{EP}L = 2m\pi$. The transition radiation is the radiation caused when an electron crosses a surface so that the dipole formed by the moving electron and its image inside the material is sharply annihilated. Although this type of radiation mostly occurs for metals, we notice that indeed it can be observed in dielectrics[52]. Moreover, when an electron traverses the edge of the material, but not directly propagating through the material, the diffraction radiation can occur due to the interaction of the electron with truncated edges. Based on our second hypothesis, the periodicity of the spatial interference fringes should be equal to the exciton-polariton wavelength of $\lambda_{EP} = 2\pi/\beta_{\text{EP}}$.

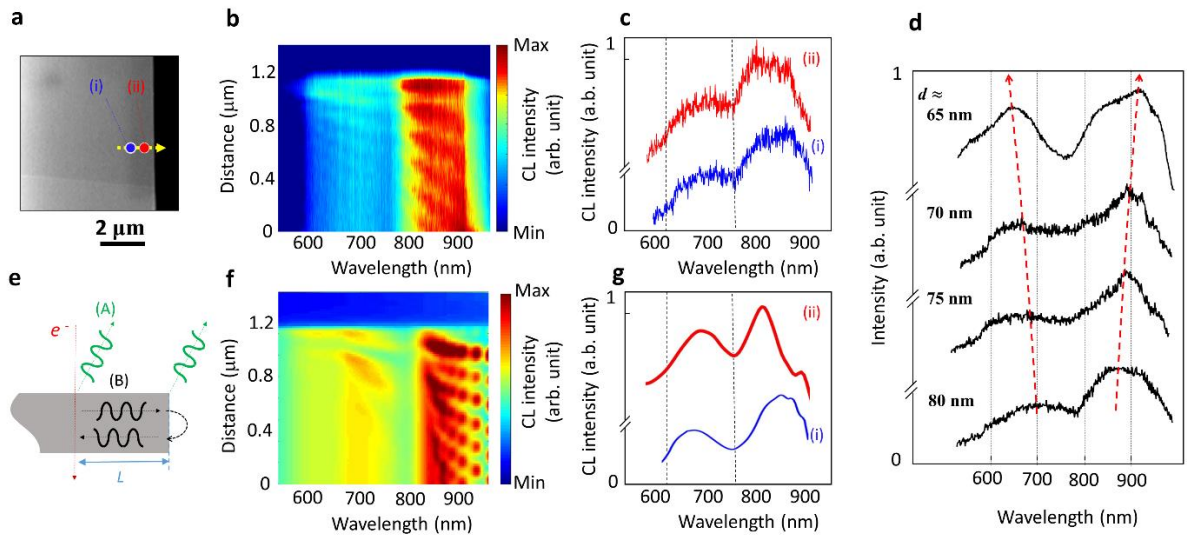

**Figure 2. CL spectroscopy of a triangular WSe$_2$ flake.** (a) SEM image of the flake. (b) CL wavelength-distance map measured along the line depicted in panel (a) showing the spatial distribution of the luminescence signal. (c) Measured CL spectra (solid lines) at selected electron excitation impact positions, as indicated in panel (a). (d) Measured CL spectra of thin films at positions away from the edge, for WSe$_2$ thin films with depicted thicknesses, showing an increase in the wavelength splitting versus thickness. (e) Schematic illustration of exciton-polariton wave reflection from the edge of the two-dimensional taper ((B); first hypothesis, see text) compared to the second hypothesis (A), originating from interference between the transition or diffraction radiation and radiation from the anomalies. Shown is the snapshot of the electron-induced $z$-component of the electric field at a given time. (f) Calculated wavelength-distance map. (g) Calculated selected spectra at positions specified in panel (a).

We provide a direct test for the proposed hypothesis, by acquiring the spatial interference fringes in a higher quality triangular flake with a thickness of 95 nm (shown in Fig. 3a and b). Our CL spectrometer is equipped with a dispersive grating, which allows for projecting the spectrally dispersed optical rays onto a CCD camera. By tilting the grating, and hence changing the central wavelength, we focus more on the UP excitonic branch, where the observed numerous interference fringes in the wavelength-distance CL map are a clear signature of the spontaneous coherence caused by the excitation of exciton-polaritons (Fig. 3b). Hyperspectral images at selected photon wavelengths capture the spatially resolved standing wave patterns, parallel to the edges (Fig. 3c). This can be understood as the constructive interference between the TR and the diffraction of the excited exciton-polaritons from the edges of the flakes at the far field. Therefore, a constructive interference pattern can be formed by the excitation of the surface exciton-polaritons with phase fronts parallel to the edges (fig. 3d). The observed peaks in the CL energy-distance maps are broad, due to the dissipative losses associated with the TE mode. The propagation constant of the slab modes can be decomposed into their real ($\beta_{\text{EP}}$) and imaginary parts ($\alpha_{\text{EP}}$), corresponding to the phase and attenuation constant, respectively (Fig. 3e). The relatively large attenuation constant for the TE mode at the measured photon energies is the reason for the large broadening of the peaks in the distance-wavelength CL map. To reveal the phase constant from the energy distance CL map, the distance between the peaks is measured at selected wavelengths as $D(\lambda_0)$ and related to the phase constant by $\beta_{\text{EP}}(\lambda_0) = 2\pi/D(\lambda_0)\cos\theta$ (see Fig. 3d for details). A direct comparison between the calculated surface exciton-polariton dispersions and the measured spectral and spatial fringes reveals a good agreement with the TM$_x$ mode, supporting our second hypothesis (Fig. 3e). This places CL spectroscopy parallel to scanning near-field optical microscopy, as a direct measure of spatially and spectrally resolved polaritonic transport mechanisms[27]. However, a better agreement can only be provided by considering edge exciton-polaritons as an individual transport mechanism parallel to surface polaritons, as will be discussed in the following.

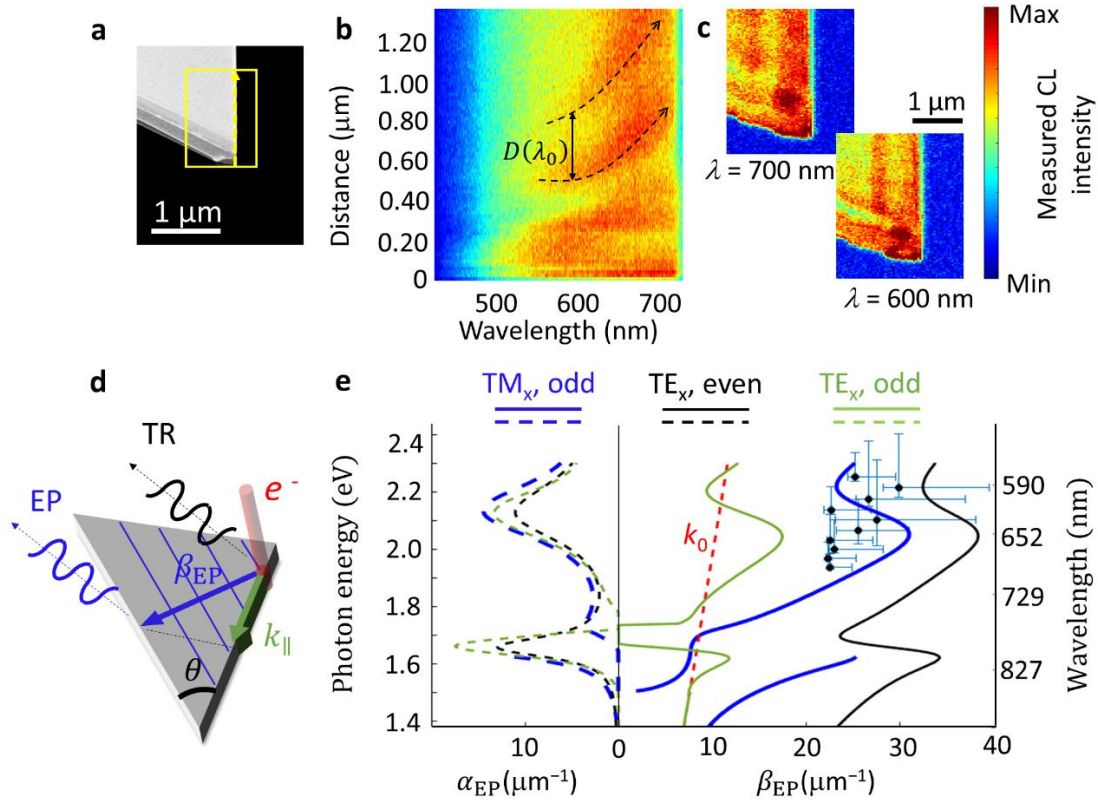

**Figure 3. Propagation mechanisms of the upper branch of the exciton-polaritons.** (a) SEM dark-field image of the investigated $WSe_2$ flake. (b) CL wavelength-distance map. Dashed arrows indicate spatial peak positions versus wavelength, where the wavelength-dependent splitting is $D(\lambda_0)$. (c) Hyperspectral images at selected wavelengths, where the interference fringes caused by the reflection from the edges of the triangle is apparent. (d) Schematic illustration of the proposed mechanism for the observed spatio-spectral interference fringes. TR: transition radiation, EP: exciton-polaritons contributing to the radiation due to their scattering from the edges. (e) Measured (scattered data) and calculated phase constant ($\beta_{EP}$; blue solid line) and attenuation constant ($\alpha_{EP}$; dashed solid line) of the surface exciton-polaritons. The dashed red line shows the optical line in free space.

**Edge exciton-polaritons.** Edge polaritons, in contrast to surface polaritons, are spatially confined to and propagate along the pristine edges of the flakes[11,32]. The spatial confinement and different screening mechanism of edge exciton-polaritons compared to bulk excitons lead to lower attenuation constants and shifted exciton energies. By scanning the edge of a $WSe_2$ flake with a trapezoidal geometry (Fig. 4a), spatial interference fringes of up to several orders are observed (Fig. 4b), due to lower attenuation constants compared to surface polaritons. The Fourier-transformed CL map allows for acquiring the dispersion of the edge exciton-polaritons (Fig. 4c). A bright intensity at $k_\parallel = 0$ is caused by the background noise, and particularly non-propagating evanescent modes with high attenuation constants. Those higher-order modes strongly contribute to the dissipation of the energy delivered by the excitation source without contributing to the desired spatial coherence associated with the propagating optical waves. In addition, however, a pronounced signal with the momenta ($\omega = k_\parallel c$) lying outside the light cone is observed. Theoretical modelling of the propagating edge exciton-polaritons are performed by calculating the optical modes supported by a rectangular rib waveguide, with the width and height of 500 nm and 100 nm, respectively. The fundamental mode of this rib waveguide has optical mode profiles confined to the edges of the waveguide (Fig. 4e). Several higher-order optical modes exist with field profiles confined to the edges and surfaces of the flakes. However, only the fundamental mode is experimentally captured, as understood by comparing the calculated dispersion diagram of the fundamental mode and the experimental results (Fig. 4c; solid lines). Hyperspectral images as well support the presumed excitation of edge exciton-polaritons and also unravel the ultra-confined mode volumes associated with these modes (Fig. 4d).

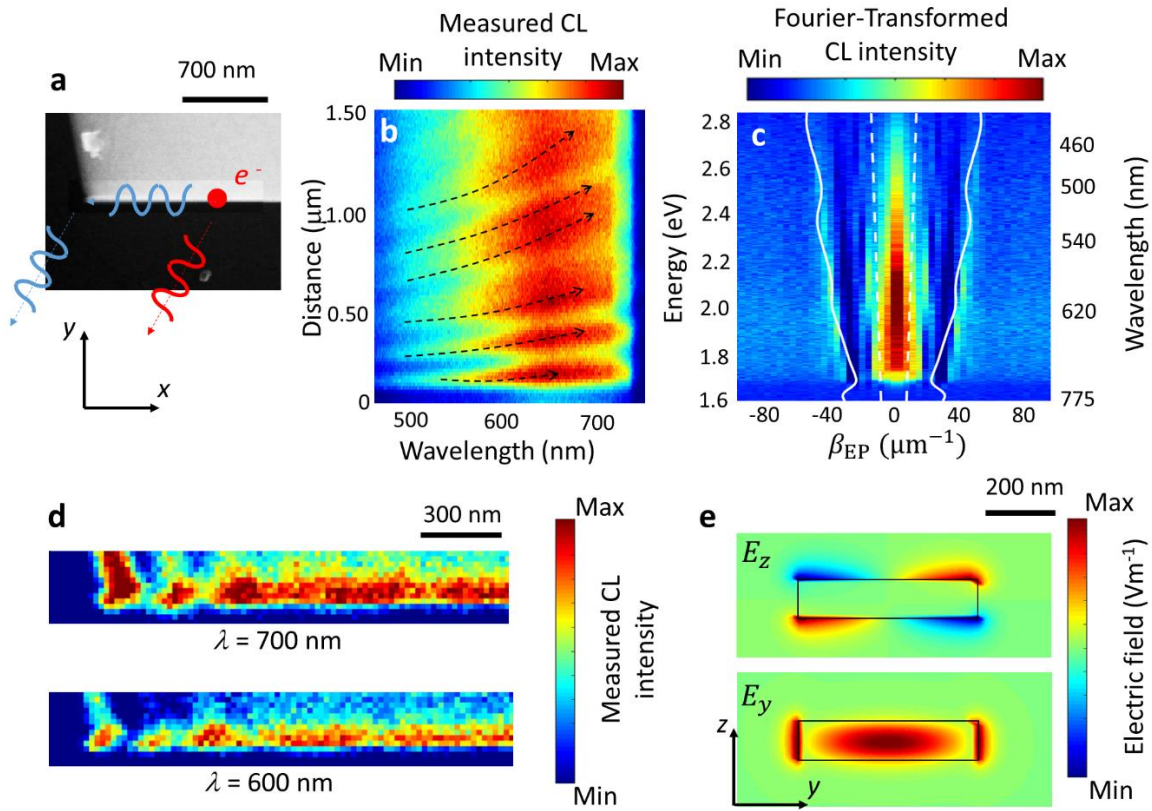

**Figure 4. Edge exciton-polaritons.** (a) The SEM dark-field image of the investigated WSe$_2$ flake. (b) Distance-wavelength CL map along the edge marked by the yellow arrow in panel (a). (c) Fourier-transformed CL map, illustrating the dispersion maps associated with the edge exciton-polaritons. Calculated dispersion diagrams of a rib WSe$_2$ waveguide with the width and heights of 500 nm and 100 nm, respectively, are shown by white and orange solid lines, for the fundamental mode. The dashed white line is the optical line in free space ($\omega = k_0 c$). (d) Hyperspectral images at selected wavelengths, where the distribution of the optical modes confined to the edge of the flakes are spatially resolved. (e) Calculated spatial distribution of the y- and z-components of the electric field at the cross section plane (yz-plane) for the fundamental guided mode propagating along the x-axis, at $E$ = 1.9 eV ($\lambda = 653$ nm).

**Photon-mediated exciton-exciton interactions.** Natural flakes with anomalies, allowing for lateral confinements of the photonic modes, provide a natural mechanism for further tuning the energy and modal configurations of the photonic excitations (see Fig. 5a). Particularly in those flakes, we indeed observe not only an omnipresent dip in the CL spectra at the wavelength associated with the A exciton, but also a series of peaks and dips in the wavelength range of the UP branch (Figs. 5b and c). These spectral fringes are revealed in numerical simulations as well (Fig. 5 d), in a good agreement with experimental results.

Higher-order photonics modes of a slab waveguide, though can cause oscillations in the LP range, they cannot result in the spectral fringes we observe in the UP range. Moreover, Cherenkov radiation (CR) caused by an electron moving inside a bulk material, can strongly interact with the excitons, similar to the photonic modes of the slab waveguide (see Supplementary Note 2 and Supplementary Fig. 4). CR can be excited and captured inside thick films with $d > 400 nm$, and can be further released to the far-field zone by virtue of its interaction with discontinuities. However, our theoretical and experimental CL analysis rules out the possibility of observing spectral fringes in the UP branch that can be related to either photonic modes or CR.

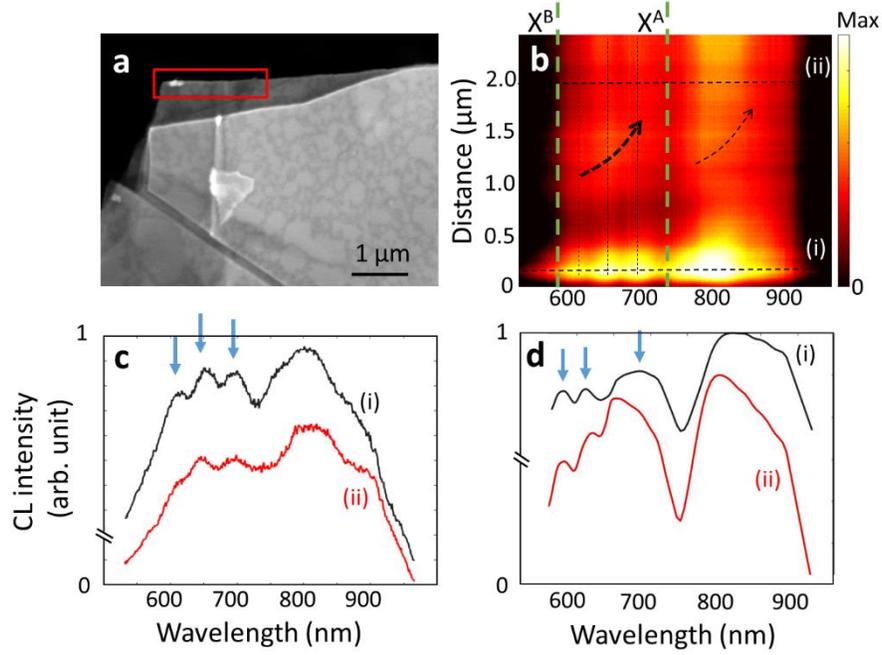

**Figure 5. Photon-mediated exciton-exciton interactions.** (a) Dark-field SEM image of the investigated flakes. (b) CL wavelength-distance map, demonstrating three finer structure peaks in the UP band, caused by three oscillator couplings (see text and Fig. 6). Measured (c) and simulated (d) spectra at selected electron impact positions indicated in panel (b).

In order to better understand the mechanisms underlying the oscillations we observe for the UP branch, we propose a model based on photon-mediated exciton-exciton interactions. Equation (1) can be modified to include photon-mediated couplings between the A and B excitons (Fig. 6a), by

$$\begin{aligned} i\dot{\vec{P}}_1 &= \left(\omega_{\text{ex,A}} - i\gamma_{\text{ex,A}}\right)\vec{P}_1 + g_1\vec{E} \\ i\dot{\vec{P}}_2 &= \left(\omega_{\text{ex,B}} - i\gamma_{\text{ex,B}}\right)\vec{P}_2 + g_2\vec{E} \\ i\dot{\vec{E}} &= \left(\omega_{\text{ph}} - i\kappa\right)\vec{E} + g_1\vec{P}_1 + g_2\vec{P}_2 \end{aligned} \quad (4)$$

where a system of three coupled harmonic damped oscillators is considered (Fig. 6b). The time-harmonic response of this coupled system of equations represents three eigenvalue solutions to the system, the values of which can be tuned by $\omega_{\text{ph}}$. We express these eigenvalues by $\omega_i$ where $i = 1, 2, 3$ (Fig. 6c). We indeed assume that excitons A and B can only indirectly interact with each other, by virtue of exchanging photons. Therefore, only two coupling coefficients between the photonic modes and the exciton peaks are considered. By considering the coupling coefficients $g_1 = 26.3$ meV and $g_2 = 65$ meV, we observe a strong shift of the second eigenwavelength from 610 nm to 730 nm, by reducing the frequency of the photonic mode from 2.2 eV to 1.6 eV (blue line, Fig. 6c). For $\omega_{\text{ph}} < \omega_{\text{ex,A}}$, we observe a strong red shift of the third eigenvalue by reducing further the photon energy (Fig. 6c, red line). By assuming a Lorentzian line shape centered at each calculated eigenvalue, we reveal an overall spectral shape that is in good agreement with the measured results represented in Figs. 5c and d. Moreover, though in general four spectral peaks are apparent in the acquired CL data (Fig. 4), we notice that the system of three coupled oscillators is sufficient to lay open the underlying mechanism, since the additional middle peak at the wavelength of $\lambda = 660$ nm can be understood by the overlapping tails of adjacent oscillators (Fig. 6d).

**Discussion**

The Cathodoluminescence response of semiconductors is naturally understood per excitation of many electron-hole pairs, where the statistical distributions of the generated photons do not resemble coherent photon statistics. We indeed ascertain that the presence of spontaneous coherence supported by the excitation of exciton-polaritons, provides evidence for CL spectroscopy to be able to probe Rabi oscillations and nonlinear exciton-photon interactions. In addition and in an inverse approach, semiconductors with room-temperature exciton excitations like TMDCs can be used to generate both X-ray photons and coherent visible-range photons upon electron irradiation, and thus provide a means for designing electron-driven photon sources of higher photon yield compared to plasmonic structures[53].

Although our results demonstrate evidence for photon-mediated exciton-exciton interactions in natural flakes, related to the energy relaxation between excitonic states, further control of the coupling mechanisms by nanostructuring and controlling the temperature will provide a toolbox for better understanding the formation of long-range spatial correlations associated with Bose-Einstein condensation.

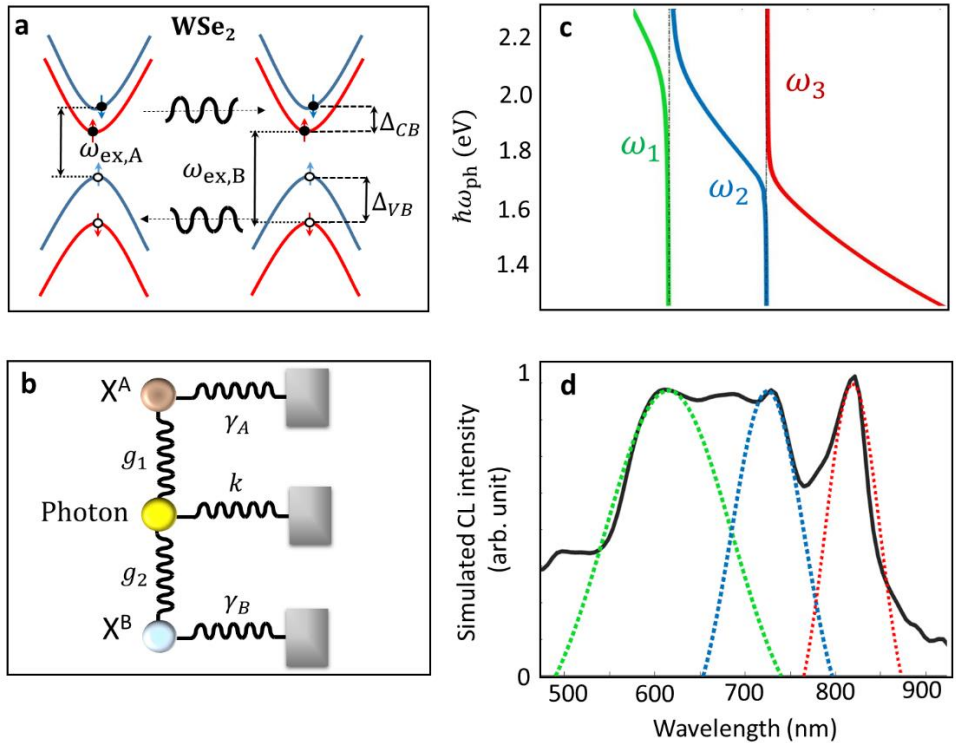

**Figure 6. Photon-mediated exciton-exciton couplings.** (a) Excitons A and B in $WSe_2$, themselves distinguished by spin-orbit interactions, can indirectly couple to each other by coherent exciton-exciton interactions. (b) A coupled three-oscillator system is introduced to describe the exciton-exciton interaction mechanism. (c) The resulting stationary solutions of the coupled three-oscillator system. (d) Three Lorentzian oscillations with the center wavelengths associated with the calculated eigenvalues result in the three-peak phenomena demonstrated in Fig. 5 in the LP polaritonic branch.

The CL response of thinner few-layered $WSe_2$ structures is more intriguing. We observe higher photon yields in general in thinner materials, due to the transition from indirect to direct bandgap[54]. Moreover, when a thin flake is transformed on top of a holy-carbon structure, the strain can heavily manipulate the spatial correlations reported so far[55] (see supplementary Fig. 3). We indeed observe competing strain-induced and exciton-polariton mechanisms in thin films of these materials.

The Stokes shift between the luminescence and absorption peaks will result in a shift of the CL peaks compared to EELS spectra[56]. However, this effect has been shown to cause only a minor effect compared to the strong-coupling effects we observe here, for nominally undoped materials, whereas it can be significantly increased for n-doped materials[57]. The short interaction of the electron beam with our structure does not cause any significant sample heating or doping, as the positions of the peaks do not depend on the acquisition time, the acceleration voltage within the range of 5 keV to 30 keV, and also not on the probe current within the range of 10 nA to 20 nA (see Supplementary Fig. 6). We therefore conclude that the possible Stokes shift has a negligible effect compared to the effect of strong exciton-photon interactions on the spectra. In summary, we have demonstrated strong coupling between excitons of $WSe_2$ and the propagating photons of freestanding thin $WSe_2$ flakes with a Rabi splitting on the order 0.24 eV. The overall energy-momentum EELS map and cathodoluminescence spectroscopy results show anticrossing of upper and lower polariton branches and long-range propagation of exciton-polaritons in high-quality atomically flat flakes. Additionally, the possibility of controlling photon-mediated exciton-exciton interactions via excited photonic modes in an optoelectronic nanocircuits is another intriguing feature of thin-film van der Waals materials. Our results thus reveal that the coupling strength of photon-mediated exciton-exciton interactions in atomically flat $WSe_2$ flakes can be further controlled by tuning the energy and field profiles of photonic modes.

## Methods

**Synthesis and exfoliation of single-crystalline $WSe_2$ crystals.** Single crystals of $2H-WSe_2$ were grown by the standard chemical vapor transport method: A near-stoichiometric mixture of high-purity W and Se with a slight Se excess (4 mg/cm$^3$) was placed in a quartz ampoule together with iodine (5 mg/cm$^3$) as transport agent; the ampoule was sealed and heated in a four-zone furnace under a temperature gradient of 920-860 °C. The samples were grown within 900 h.

Thin nanosheets were prepared by applying liquid phase exfoliation (LPE) from the bulk $WSe_2$ in isopropanol (Merck, ≥99.8%) (Supplementary Fig. 5). Exfoliation was performed with an ultrasonicator (320 W, Bandelin Sonorex, RK100H) equipped with a timer and heat controller to avoid solvent evaporation. Sonication was conducted in an ice bath by applying a cycle program of 5 min on, followed by 1 min off for a total duration of 60 min. The resulting suspension was drop casted either on a high purity graphite planchet or holey carbon mesh grid for further characterization.

**Cathodoluminescence imaging.** High-resolution SEM observations were performed using an optical field emission microscope (Zeiss SIGMA) operated at 30 kV. To perform the CL measurements, a beam current of 14 nA was utilized to scan the specimen and excite the surface. The generated CL radiation was collected and analysed using a CL detector (Delmic B.V) equipped with an off-axis silver parabolic mirror (focal distance: 0.5 mm). The acceptance angle of the mirror and dwell time were 1.46π sr and 400 μs, respectively. The mirror was positioned on the upper side of the specimen and a hole with a diameter of 600 μm above the focal point was supporting the electron excitation.

**Numerical simulations.** We have employed both a home-built finite-difference time domain method and the COMSOL Multiphysics software to gain insight into the temporal distribution of the electron-induced radiation and exciton-photon interactions in our $WSe_2$ sample. Our results show good agreement with each other and with the experimental results.

To simulate the CL spectra, we have employed a classical method to model an electron beam by a current density distribution corresponding to a swift electron charge at kinetic energy of $E$ =30 keV[58,59]. The mesh size is 2 nm and a higher-order absorbing boundary conditions is used. The dielectric function of $WSe_2$ was modeled by the two critical point functions demonstrated in the supporting information. The CL spectra are calculated by employing the discrete Fourier transformation on the field distributions at planes circumventing the structure at distances of 1 μm from the structure, and calculating the Poynting vector. The convergence was achieved by employing 8,000 time iterations.

To calculate the EELS and CL, in the frequency domain, we have employed the radiofrequency (RF) toolbox of COMSOL in a 3D simulation domain, where the Maxwell equations are solved in real space and in the frequency domain. We have utilized an oscillating "Edge Current" as a fast electron beam along a straight line representing the electron beam. The current was expressed by $I = I_0 \exp(i\omega z/v_e)$. The CL was calculated by using boundary probe at a plane normal to the electron beam. The dielectric function of $WSe_2$ was defined using the interpolation function of COMSOL.

## Acknowledgements

This project has received funding from the European Research Council (ERC) under the European Union's Horizon 2020 research and innovation programme, Grant Agreement No. 802130 (Kiel, NanoBeam) and Grant Agreement No. 101017720 (EBEAM). Financial support from Deutsche Forschungsgemeinschaft under the Art. 91 b GG Grant Agreement No. 447330010 and Grant Agreement No. 440395346 is acknowledged.

## Data availability

The data that support the findings of this study are available from the corresponding author upon reasonable request.

## Author Contributions

MT has performed the CL measurements and data analysis, and fabricated the thin-film samples. FD and NT have performed the simulations. FKD and KR have provided the bulk crystals. NT and KR have written the manuscript. All the coauthors have contributed to discussions. NT has supervised the work.

## Ethics Declarations

### Competing interests

The authors declare no competing interests.